\documentclass[letter, 11pt]{revtex4-1}
\usepackage{epsfig,epstopdf,amsmath,amssymb,graphicx,subfigure,verbatim,natbib}

\setcitestyle{super}

\usepackage{calligra}
\DeclareMathAlphabet{\mathcalligra}{T1}{calligra}{m}{n}
\DeclareFontShape{T1}{calligra}{m}{n}{<->s*[2.2]callig15}{}

\begin{document}

\author{John A. Brehm}
\email{brehmj@sas.upenn.edu}
\affiliation{The Makineni Theoretical Laboratories, Department of Chemistry, University of Pennsylvania, 231 S. 34th Street, Philadelphia, Pennsylvania  19104-6323, USA}

\author{Hiroyuki Takenaka}
\email{hitak@sas.upenn.edu}
\affiliation{The Makineni Theoretical Laboratories, Department of Chemistry, University of Pennsylvania, 231 S. 34th Street, Philadelphia, Pennsylvania 19104-6323, USA}

\author{Chan-Woo Lee}
\email{leechanw@sas.upenn.edu}
\affiliation{The Makineni Theoretical Laboratories, Department of Chemistry, University of Pennsylvania, 231 S. 34th Street, Philadelphia, Pennsylvania 19104-6323, USA}

\author{Ilya Grinberg}
\email{ilya2@sas.upenn.edu}
\affiliation{The Makineni Theoretical Laboratories, Department of Chemistry, University of Pennsylvania, 231 S. 34th Street, Philadelphia, Pennsylvania 19104-6323, USA} 
\author{Joseph W. Bennett}
\email{bennett@physics.rutgers.edu}
\affiliation{The Makineni Theoretical Laboratories, Department of Chemistry, University of Pennsylvania, 231 S. 34th Street, Philadelphia, Pennsylvania 19104-6323, USA}

\author{Michael Rutenberg Schoenberg}
\email{michael.rutenbergschoenberg@yale.edu}
\affiliation{The Makineni Theoretical Laboratories, Department of Chemistry, University of Pennsylvania, 231 S. 34th Street, Philadelphia, Pennsylvania 19104-6323, USA}

\author{Andrew M. Rappe}
\email{rappe@sas.upenn.edu}
\affiliation{The Makineni Theoretical Laboratories, Department of Chemistry, University of Pennsylvania, 231 S. 34th Street, Philadelphia, Pennsylvania 19104-6323, USA}

\date{\today}

\begin{abstract}
Using density functional theory (DFT) within the local density approximation (LDA), we calculate the physical and electronic properties of PbTiO$_3$ (PTO) and a series of hypothetical compounds   PbTiO$_\text{3-$x$}$S$_\text{$x$}$ $x$ = 0.2, 0.25, 0.33, 0.5, 1, 2, and 3 arranged in the corner-sharing cubic perovskite structure. We determine that replacing the apical oxygen atom in the PTO tetragonal unit cell with a sulfur atom reduces the $x$ = 0 LDA calculated band gap of 1.47 eV to  0.43 - 0.67 eV  for $x$ = 0.2 - 1 and increases the polarization.  PBE0 and $GW$ methods predict that the compositions $x$ = 0.2-2 will have band gaps in the visible range. For all values of $x$ {\textless } 2, the oxysulfide perovskite retains the tetragonal phase of PbTiO$_\text{3}$, and the $a$ lattice parameter  remains within 2.5\% of the oxide.  Thermodynamic analysis indicates that chemical routes using high temperature gas, such as H$_2$S and CS$_2$, can be used to substitute O for S in PTO for the compositions $x$ = 0.2 - 0.5.
\end{abstract}

\title{Density Functional Theory Study Of Hypothetical PbTiO$_\text{3}$-Based Oxysulfides} 
\maketitle

\section{Introduction}

In this paper, we use first principles  calculations to study the physical  and electronic properties of hypothetical polar oxysulfide perovskite solid solutions of the chemical formula PbTiO$_\text{3-$x$}$S$_\text{$x$}$.  We seek to identify new photovoltaic materials for efficient solar energy conversion. Photovoltaic  materials must have band gaps in the range 1.1 - 2 eV to provide strong light absorption and energy conversion.  The best single-junction materials, such as silicon, CdTe, and copper indium selenide, exhibit gaps near the Shockley and Queisser (SQ) detailed balance model optimal value of 1.3~eV.
Photons with energies less than the band gap will not promote electrons to the conduction band, while electrons absorbing photons with energies greater than the conduction band minimum will lose energy as the electrons decay to the band edge. 
To surpass the single-material SQ limit,  photovoltaics with a range of band gaps are arranged in multi-junction solar cells, $e.$ $g.$  CuIn$_\text{$x$}$Ga$_\text{1-$x$}$Se$_\text{2}$ and InGaP/GaAs/InGaAsN/Ge.\cite{Teshima07p1012-Y03-19,Ohsita11p328} 
In addition to a good match between the band gap and the solar spectrum, excited carrier recombination must be prevented in order to obtain the photocurrent.  This is typically done by an electric field generated at a $p-n$ junction that moves the holes and the excited electrons in opposite directions.
Another recently studied method has been to use ferroelectric materials, for which the strong inversion symmetry breaking and spontaneous polarization give rise to the separation of charge carriers in the bulk of the material (bulk photovoltaic effect). The known perovskite BiFeO$_3$\cite{Basu08p091905} and new materials including [KNbO$_3$]$_{1-x}$[BaNi$_{1/2}$Nb$_{1/2}$O$_{3-\delta}$]$_x$,\cite{Grinberg13p509} KBiFe$_2$O$_5$,\cite{Zhang13p1265} and Bi$_4$Ti$_3$O$_{12}$-LaCoO$_3$\cite{Choi12p1038}  are examples of polar oxide materials with band gaps in the visible range in which the bulk photovoltaic effect has been observed.

We choose the classic PbTiO$_3$ (PTO) ferroelectric perovskite oxide as a basis for alloying with sulfur on the oxygen sites.  PTO is highly polar ($P$ = 0.88 C/m$^2$) and  has an indirect band gap of \text{$\approx$}3.45 - 3.6 eV.\cite{Peng92p21,Moret02p468, Bao02p269,Grinberg04p220101} Excitation across the gap is essentially a charge transfer from the O 2$p$ orbitals to the Ti 3$d$ orbitals.  Therefore, substitution of the more electronegative oxygen by the less electronegative sulfur should lead to a lower band gap, while possibly preserving the ferroelectricity.   The substitution of oxygen by sulfur or vice versa in order to tailor band gaps is well documented in other materials.\cite{Kosugi98p273, Barreau02p439, Ogisu08p11978} Unlike replacement of oxygen with nitrogen or titanium with nickel to lower band gaps,\cite{Asahi01p269, Bennett08p17409}  the substitution of isovalent sulfur in place of oxygen does not require vacancies to preserve charge neutrality.

While simple corner-sharing oxysulfide
perovskites have yet to be reported in the literature, closely related
and more complicated oxysulfides do exist. They have been reported as either
Ruddlesden-Popper phases\cite{Ishikawa04p2637, Rutt03p7906, Charkin10p2012} or as layered materials with perovskite-like
oxide layers alternating with either antifluorite or rock salt sulfide
layers.\cite{Zhu00p26, Hyett08p559, Meignen04p2810, Smura11p2691} Thus the current work is novel in that it explores the feasibility of synthesizing a purely corner-sharing perovskite phase.

\section{Methodology}

All density functional theory (DFT) calculations in this study are performed using the local density approximation (LDA). The DFT packages we use in this study are ABINIT\cite{Gonze02p478}  and Quantum Espresso.\cite{Giannozzi09p395502}  The atoms are represented by norm-conserving optimized pseudopotentials \cite{Rappe90p1227} generated using OPIUM, \cite{OpiumSourceforgenet11p1} and all, except oxygen, are further refined using the designed non-local methodology. \cite{Ramer99p12471} We pseudize the following orbitals: $2s$ and $2p$ for O; $3s$, $3p$, and $3d$ for S; $5d$, $6s$, and $6p$ for Pb; and $3s$, $3p$, $3d$, $4s$, and $4p$ for Ti.  The pseudopotentials are optimized for a 50 Ry plane-wave cutoff, and all solid-state calculations use this value.  

ABINIT is used for relaxation calculations, in order to determine unit cell parameters, atomic positions, and relative energies. For the standard perovskite unit cell of five atoms, a Monkhorst-Pack (MP) k-point grid\cite{1976Monkhorst5188} of 8$\times$8$\times$8 is used (though results using a 4$\times$4$\times$4 grid are quite similar to those using an 8$\times$8$\times$8 grid). For calculations requiring a doubling of the unit cell in a Cartesian direction, the MP grid is set to 4 for that dimension.   

In order to determine the preferred location of S atoms for  PbTiO$_\text{3-$x$}$S$_\text{$x$}$ with $x$ = 1 and $x$ = 2, we perform two sets of calculations.  In the first set, a five-atom unit cell is used.  We evaluate all the possible locations of a minority species anion in a tetragonal perovskite cell.  We also consider the paraelectric, cubic perovskite and displace the anions above and below an imaginary center plane that intersects four of the six anions of the octahedron.   For initial cell parameters, we use two strain states as well:  a compressed case and an expanded case.   For the compressed case, the cell parameters are $a$ = 3.87 {\AA } and $c$ = 4.07 {\AA } for the tetragonal cells  and $a$ = 4.09 {\AA } for the cubic cells.  (As will be shown later, the cell parameters used  for the tetragonal/compressed case were calculated from a five-atom PTO relaxation.)  The $a$ lattice parameter in the cubic case is determined by setting the Pb-O-Pb face diagonal length equal to twice the sum of the  ionic radii of  a 12-coordinated Pb$^\text{2+}$ (1.49 {\AA}) and a six-coordinated O$^\text{2-}$ (1.40 {\AA}). \cite{2011Seshadri1}  For the expanded cells, cell parameters are increased by a factor of 1.31, which represents the radius ratio of six-coordinated S$^\text{2-}$ (1.84 \AA) to six-coordinated O$^\text{2-}$.\cite{2011Seshadri1}  

For the second set of calculations, we consider ten-atom unit cells consisting of two PTO unit cells with either the $a$ or the $c$ lattice parameter doubled.  Ten-atom unit cells also accommodate $x$ = 0.5 concentrations.  We evaluate all of the cases in which a S atom may occupy any of the six O positions for the $x$ = 0.5 system, and all of the ways that two minority species anions may occupy the six anion locations for the $x$ = 1 and $x$ = 2 systems.     The systems $x$ = 0 and $x$ = 3, with only one anion species, have only one configuration and were relaxed from five-atom cubic unit cells with an initial starting $a$ lattice parameter derived from ionic radii sums.  

In order to extend our analysis to lower concentrations of S,  we create unit cells by inserting one, two and three PTO layers into the relaxed  $x$ = 0.5 structure while keeping the S atom confined to the (001).  These compositions have $x$ = 0.33, 0.25, and 0.2 respectively.  All systems are considered to be fully relaxed when successive self-consistent iterations yield total energy differences less than 10$^\text{-8}$ Ha/cell and atomic forces less than 10$^\text{-4}$ Ha/Bohr.  The FINDSYM package\cite{2011Stokes1,Stokes05p237} is used to determine the space groups of the relaxed structures.  In order to assess whether or not any of the systems prefer a Glazer tilt system structure,\cite{Glazer72p3384} a set of relaxation calculations is performed on 40-atom 2$\times$2$\times$2 unit cells for the $x$ = 0, 0.5, 1, 2, and 3 compositions.  The starting locations of the S atoms for $x$ = 0.5, 1, and 2 are determined from the earlier five- and ten-atom relaxations.  Polarization calculations are carried out using ABINIT,  while band gap and projected density of states (PDOS) calculations are performed using Quantum Espresso with 12$\times$12$\times$12 k-point grids. Post-DFT band gap investigation is carried out using the PBE0\cite{Perdew96p9982} method as implemented in Quantum Espresso, and the $GW$\cite{Hedin65pA796} method as implemented in ABINIT.\cite{Lebegue03p155208,Bruneval06p045102, Anisimov00}  The nature of the bonding in $x$ = 0, 0.5, 1, and 2 systems  is assessed using a  Bader charge  analysis package.\cite{Henkelman06p254, Henkelman13p1}

\section{Results and Discussion}

\subsection {Structural and electronic results for end-members PbTiO$_\text{3}$ and PbTiS$_\text{3}$.}

Our  calculated results for the end member,  PbTiO$_\text{3}$,  are in agreement  with earlier published theoretical work.\cite{Piskunov03p165, Meyer99p395, SaghiSzabo99p12771}  The relaxed lattice parameters are  $a$ = 3.87 \text{\AA } and $c$ = 4.07 \AA, yielding a $c/a$ ratio of 1.05.  We find an indirect (X-$\Gamma$) band gap of 1.47 eV and polarization (strictly in the $z$ direction) of 0.85 C/m$^\text{2}$.   These results, as well as all the physical and electronic properties for PbTiO$_\text{3-$x$}$S$_\text{$x$}$ $x$ = 0 and $x$ = 3 and high symmetry PbTiO$_\text{3-$x$}$S$_\text{$x$}$ $x$ = 0.2 - 2 are listed in Tables \ref{elecres} and \ref{elecres2}.   Since the LDA calculated band gap underestimates the experimental band gap, post-DFT methods must be applied.  For PTO, we calculate a band gap of 3.44 eV using the PBE0 method, (with the $\alpha$ parameter set to 0.25), and a gap of 4.10 eV by the $GW$ method.  To our knowledge, only one other theoretical study\cite{Gou11p205115} has applied the $GW$ method to calculate the band gap in PTO and the calculations are consistent.  Since the $GW$ result overestimates the experimental PTO band gap, we investigate whether the semi-core states incur greater exchange-correlation errors by calculating the $GW$ gap without the semi-core states of Pb and Ti in the valence space.  However, this leads to only a 0.2 eV reduction in the PTO band gap.   We therefore surmise that the PBE0 method is probably a slightly better predictor of the  band gap for the oxysulfide systems.  We do not use the LDA+$U$ method, as an earlier paper reports a band gap of only 2.17 eV with this technique.\cite{Gou11p205115}

Within the cubic corner-sharing motif, PbTiS$_\text{3}$ is found to be a  $a$$^\text{+}$$b$$^\text{-}$$b$$^\text{-}$ tilt system with $a$ \text{$\approx$ }12$^\circ$ and $b$ \text{$\approx$ }15$^\circ$, as determined from the 40-atom 2$\times$2$\times$2  relaxation. A further relaxation, this time on a standard 20-atom unit cell arranged in the $Pnma$ structure,  yields the same structure as the 40-atom relaxation.  Thus, we conclude that the ground state for PbTiS$_\text{3}$ arranged in the corner-sharing perovskite mode is $Pnma$.   While the LDA electronic structure calculations predict that it is metallic and nonpolar, the PBE0 calculated gap is 0.86 eV.    With the minimum representation for the $Pnma$ tilt system being a 20-atom unit cell, the $GW$ method is not applied here as it is computationally expensive.  \cite{Gou11p205115}  Also,  if we restrict the five atom unit cell to the $P4mm$ space group, which is higher in relative energy to the $Pnma$ phase, even though an LDA band gap calculation again shows the compound to be metallic, a gap of 1.60 eV is obtained by the PBE0 method and a 1.19 eV gap by the $GW$ method.  With a band gap in the visible range and associated polarization stemming from its non-centrosymmetric nature, PbTiS$_\text{3}$ arranged in the $P4mm$ space group has the properties required for a bulk photovoltaic effect material.  This contrasts with the low energy $Pnma$ formation, which is centrosymmetric and thus not suitable for bulk photovoltaic effect purposes.  However, tempering these results is the fact that PbTiS$_\text{3}$ has not been made, and the only synthesized stoichiometries known for the Pb-Ti-S system are misfit layered compounds with chemical formula (PbS)$_\text{1.18}$(TiS$_\text{2}$)  and  (PbS)$_\text{1.18}$(TiS$_\text{2}$)$_\text{0.8475}$.\cite{Meershcaut92p129}  In these compounds, distorted rock salt-like PbS layers are intercalated with TiS$_\text{2}$ edge-sharing sheets.

\begin{table*}[h]
\caption{\label{elecres}  Calculated structural properties for PbTiO$_\text{3}$,  PbTiS$_\text{3}$, and high symmetry PbTiO$_\text{3-$x$}$S$_\text{$x$}$ $x$ = 0.2 - 2. The $c/a$ ratios are normalized according to the number of octahedra in the unit cell. }
\begin{ruledtabular}
\begin{tabular}{c|cccc|c} 
 & \multicolumn{3}{c} {Unit Cell Lengths (\AA)} &  &  \\
 & \multicolumn{3}{c} { (and Unit Cell Angles where $\not=$ 90$^\circ$)} &$c/a$ &Space  \\

$x$ & $a$ & $b$ &$c$ & Ratio & Group\\ \hline
0 & 3.87 & 3.87 &4.07 & 1.05 & $P4mm$ \\
0.2 & 3.86 & 3.86 &21.70 & 1.12 &  $P4mm$\\
0.25 & 3.85 & 3.85 &17.65 & 1.15 & $P4mm$\\
0.33 & 3.85 & 3.85 &13.57 & 1.17 & $P4mm$\\
0.5 & 3.84 & 3.84 &9.56  & 1.24  & $P4mm$ \\
1 & 3.78 & 3.78 &5.64 & 1.49 & $P4mm$ \\ \hline
2 & 4.86 ($\alpha$ = 90.37$^\circ$) &4.86 ($\beta$ = 89.63$^\circ$) &3.76 ($\gamma$=90.23$^\circ$)& 0.77 & $P1$ \\ \hline
3 & 9.31& 9.31 & 9.25 & 0.99 & $Pnma$ \\
\end{tabular}
\end{ruledtabular}
\end{table*}

\begin{table*}[h]
\caption{\label{elecres2} Calculated electronic properties for PbTiO$_\text{3}$,  PbTiS$_\text{3}$, and high symmetry PbTiO$_\text{3-$x$}$S$_\text{$x$}$ $x$ = 0.2 - 2.  For $x$ = 0.5, 1, and 2, the  LDA band gaps and total polarization ranges for all ten-atom unit cells are included in parentheses.  $GW$ band gaps calculated using pseudopotentials without semi-core states are listed in parentheses.  In comparison, the experimental band gap for $x$ = 0 has been reported between \text{$\approx$}3.45 - 3.6 eV.\cite{Peng92p21,Moret02p468, Bao02p269} }
\begin{ruledtabular}
\begin{tabular}{c|ccc|ccccc} 
 & \multicolumn{3}{c|} {Band Gap (eV)} &   \multicolumn{4}{c} {Polarization (C/m$^\text{2}$)}   \\
$x$ & LDA & PBE0 & $GW$ & $P$$_\text{$x$}$  & $P$$_\text{$y$}$   & $P$$_\text{$z$}$   & $P$\\ \hline
0 & 1.47 & 3.44 & 4.10 (3.90) & - & - & 0.85 & 0.85 \\
0.2 & 0.45 & 2.08 & & - & - & 0.87 & 0.87\\
0.25 & 0.46 & 2.11 & & - & - & 0.87 & 0.87\\
0.33 & 0.48 & 2.12 && - & - & 0.88 & 0.88 \\
0.5 & 0.60 (0.60 - 0.89) & 2.19 & 2.69 (2.50) & - & - & 0.92 & 0.92 (0.92 - 1.13)\\
1 & 0.67 (0.14 -1.15) & 2.25 & 2.60 (2.41) & - & - & 1.11 & 1.11 (0.71 - 1.13)\\
2 & 0.33 (0.05 - 1.18) & 1.77 & 1.38 (1.33) & 0.24 & 0.23 & 0.27 & 0.43 (0.28 - 0.95)\\
3 & 0.00 & 0.86 & 1.19$^a$ & -&  -  & - & - 
\end{tabular}
\end{ruledtabular}
$^a$Band gap evaluated for a five-atom unit cell arranged in the $P4mm$ space group.
\end{table*}

\subsection {Structural Properties of  PbTiO$_\text{3-$x$}$S$_\text{$x$}$.}
We find that S substitution $x$ = 0.5 and 1 preserves the tetragonal  $P4mm$ structure and leads to a small decrease of the $a$ lattice constant and  a considerable increase in the $c$ lattice constant and the $c$/$a$ ratio, reaching $c$ = 5.64~\AA\ and $c$/$a$ = 1.49 for $x$ = 1.  (See Table \ref{elecres}.) For $x$ = 0.5,  the two octahedra differ, with one having $c$ = 4.58 \text{\AA } and the other having $c$ = 4.98 \text{\AA}, as a result of different chemical environments.
 The large tetragonality values are in  agreement with octahedral cage sizes that have been observed  experimentally in oxysulfides.   For example, in $A$$_\text{2}$CoO$_\text{2}$Cu$_\text{2}$S$_\text{2}$ ($A$= Sr, Ba) solid solutions,  Smura \emph{et al}.\cite{Smura11p2691} have found $c/a$ ratios ranging between 1.52 and 1.66 for CoO$_\text{4}$S$_\text{2}$ octahedra.  Similarly, Ishikawa \emph{et al}.\cite{Ishikawa04p2637} have found an average $c/a$ ratio of 1.6  for $Ln$$_\text{2}$Ti$_\text{2}$O$_\text{5}$S$_\text{2}$  ($Ln$ = Pr, Nd, Sm, Gd, Tb, Dy, Ho, Er).     

\begin{figure}[h]
\vspace{0pt}
\includegraphics[scale = 0.40]{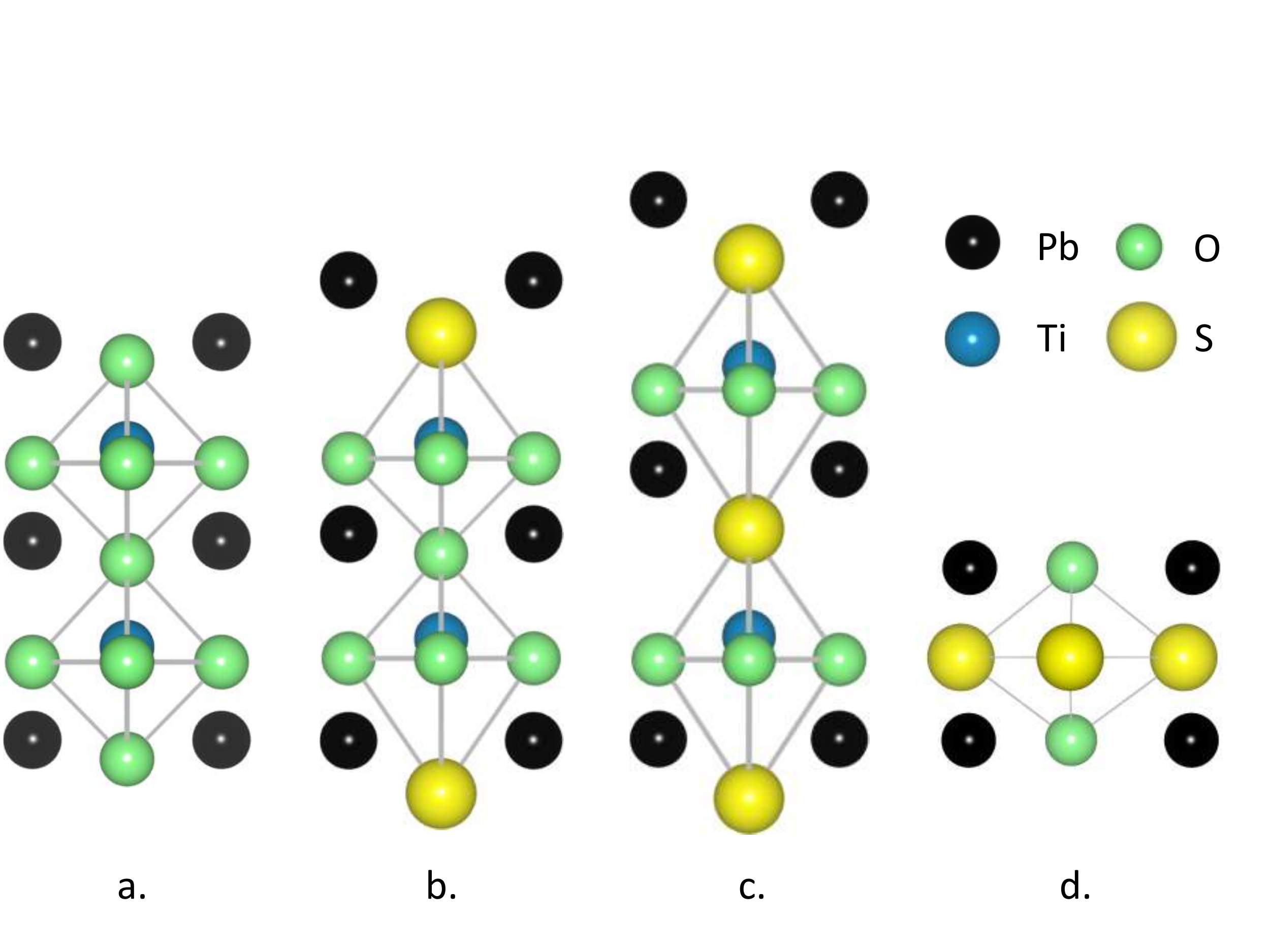}
\vspace{-20pt}
\caption{\label{polar} Relaxed structures of  PbTiO$_\text{3-$x$}$S$_\text{$x$}$: (a.) two unit cells of $x$ = 0, (b.) one unit cell of $x$ = 0.5, (c.)  two unit cells of $x$ = 1, and (d.) one unit cell of $x$ = 2.  The view is of the $ac$-plane. All dimensions and ionic radii  are to scale, except the Ti ions which are enlarged for clarity.  Images created with VESTA.\cite{VESTA}}
\end{figure}

The preferred location for the substituent S atoms is apical for $x$ = 0.5 and 1, such that the $B$-S-$B$ bonds are along the $c$-axis.  This location for the S atom in $x$ = 1 and $x$  = 0.5 has been observed experimentally in layered oxysulfide perovskites where the S-$M$-S bonds in $M$O$_\text{4}$S$_\text{2}$ single octahedra\cite{Smura11p2691} and the S-$M$-O-$M$-S bonds for pairs of octahedra of the form $M$O$_\text{5}$S connected by an oxygen,\cite{Ishikawa04p2637}  are linear along the $c$ axes of elongated octahedra.    We do not observe any octahedral tilting for these structures.   The structures for $x$ = 0, 0.5, 1, and 2 are shown in Figure \ref{polar}.   

Relaxation calculations on compositions $x$  = 0.2, 0.25, and 0.33 show that the $a$ lattice parameters are within 0.5\% of each other and the $x$ = 0 and $x$ = 0.5 compositions,  the $c$ lattice parameter increases by multiples of the length of the PTO unit cell, 4.07 \text{\AA}, as one would expect, and the high symmetry $P4mm$ phase is maintained.

The  PbTiO$_\text{3-$x$}$S$_\text{$x$}$ compositions $x$ = 0 - 1 have very similar cation-anion bond lengths.  These are reported in Table \ref{bondlens}.  Indeed, all Ti-O lengths are nearly equal to the PTO values.  In order to corroborate our Pb-S bond length results, we have also calculated the lattice constant and Pb-S bond length for rock salt PbS.  Our PbS lattice constant of 5.85 \text{\AA } is in excellent agreement with the previously reported theoretically calculated values.\cite{Lach-hab02p833,Dantas08p1451}  Also, our Pb-S bond length of 2.93 \text{\AA } is within 0.02 \text{\AA } of the values reported in these works.  Thus, the compositions PbTiO$_\text{3-$x$}$S$_\text{$x$}$ with $x$ = 0 - 1 form crystals that maintain the nearest neighbor (NN) cation-anion bond lengths of PTO and rock salt PbS.

\begin{table*}
\caption{\label{bondlens} Selected cation-anion bond lengths and $z$-displacements for PbTiO$_\text{3-$x$}$S$_\text{$x$}$ $x$ = 0 - 1.  All lengths in \text{\AA}.  NN = nearest neighbor.  For $x$ = 0.20, 0.25, and 0.33, average values for Pb - NN apical O, Ti - equatorial O, and Ti - NN apical O are tabulated.  Pb - S and Ti - NN S average values are also listed.  $\Delta$$z$ is defined as the separation in the $z$ coordinate between two ions.}
\begin{ruledtabular}
\begin{tabular}{c|cc|cc|cc|c|c} 
      & \multicolumn{2}{c|} {Pb - NN apical O}   & \multicolumn{2}{c|} {Pb-S} & \multicolumn{2}{c|}{Ti - equatorial O} & Ti - NN apical O &Ti - NN S \\
 $x$& bond length & $\Delta$$z$       & bond length & $\Delta$$z$    & bond length & $\Delta$$z$ & bond length & bond length \\     \hline
0 & 2.76 & 0.39 & - & - & 1.96 & 0.30 & 1.78 & - \\
0.20 & 2.76&0.40 & 2.92 & 1.05 &1.95 &0.32 &1.78 & 2.31 \\
0.25 &2.75 &0.40 & 2.92 & 1.06 &1.95 &0.32 &1.78 & 2.31 \\
0.33 & 2.75&0.40 & 2.92 &  1.06 &1.95 &0.33 &1.78 & 2.31 \\
0.50 & 2.75 & 0.41 & 2.93 & 1.10 & 1.95 & 0.35 & 1.77 & 2.30 \\
1 &     - & - & 2.95 & 1.24 & 1.95 & 0.48 & - & 2.25 \\
\end{tabular}
\end{ruledtabular}
\end{table*}

The results for the $x$ = 2 composition  differ markedly from the $x$ = 0.5 and $x$ = 1 systems.  The $x$ = 2 composition has very low symmetry ($P1$), $c/a$ {\textless } 1, and lattice angles that all differ from 90$^\circ$ by up to 0.37$^\circ$.  Relaxation of the various  ten-atom unit cells shows  the minority species anions, in this case O, again prefer to be located trans to each other, forming \text{$\approx$}180$^\circ$   O-Ti-O angles.  Relaxation of the 40-atom 2$\times$2$\times$2 unit cell does not result in any tilting.
 

Further analysis of the atom locations in the $x$ = 2 unit cell shows significant distinctions from the $x$ = 0 - 1 compounds. First, the anion displacement is no longer only in the $z$ direction.  The S anions are displaced  from high-symmetry positions up to  \text{$\approx$}0.08 \text{\AA } in the $x$ and $y$ directions, and the O anion is displaced 0.04 \text{\AA } in both $x$ and $y$.  Moreover, while the short Ti - apical O  bond length is 1.75 \text{\AA }, which is almost identical to that for compositions $x$ = 0 - 1, (see Table 3),  the displacement of the Ti atom relative to the equatorial anions  is much reduced relative to those other compositions.  The $x$ = 0 - 1 compositions have displacements strictly in the $z$ direction with magnitudes monotonically increasing from 0.30  to 0.48~\text{\AA}; however the $x$ = 2 system shows displacements in all three Cartesian directions:  0.08 \text{\AA } in $x$, 0.02~\text{\AA } in $y$, and  0.20~\text{\AA } in $z$.  This leads to a total displacement  magnitude 0.22 \text{\AA }.    Unlike the $x$ = 0 - 1 compounds, the Pb sublattice only displaces by a relatively smaller 0.07~\text{\AA } in the $z$ direction, but it also displaces 0.19 \text{\AA }  in the $x$  and $y$ directions.  The total displacement of 0.27 \text{\AA } is smaller than the 0.39 \text{\AA} Pb - O $z$ displacement in PTO and much smaller than the  Pb - S displacements in the compounds with $x$ = 0.2 - 1 (which are greater than 1~\text{\AA}).   The offset of Pb in the $xy$ plane leads to two distinct Pb - S bond lengths:  one that averages 2.90~\text{\AA}, and one much larger, averaging 3.21~\text{\AA}.  The smaller Pb - S bond length is in agreement with the data for the $x$ = 0.2  - 1 compounds, and is only 0.03 \text{\AA } smaller than the calculated Pb - S bond length for rock salt PbS.  The Ti - S bond lengths range from 2.40 \text{\AA } to 2.48 \text{\AA }.  These bond lengths are basically the sum of the ionic radii of Ti  (0.605 \text{\AA }) and S  (1.84 \text{\AA}).

\subsection {Electronic properties of PbTiO$_\text{3-$x$}$S$_\text{$x$}$.}

Table \ref{elecres2} clearly shows that as $x$ increases from 0 to 1, $P$ = $P_z$, and $P$ increases monotonically from 0.85 to 1.11 C/m$^\text{2}$.  For $x$ = 2, $P$ is reduced to 0.43 C/m$^\text{2}$, with significant $P$ components in each Cartesian direction of \text{$\approx$}0.25 C/m$^\text{2}$.  This suggests a morphotropic phase boundary between $x$ = 1 and $x$ = 2, potentially leading to high piezoelectricity.  The reduced $P$ for $x$ = 2, with significant components in all three Cartesian directions, is confirmed by the smaller displacement vectors for Pb and Ti as described earlier.

\begin{figure*}[h]
\vspace{-40pt}
\includegraphics[scale=1.4]{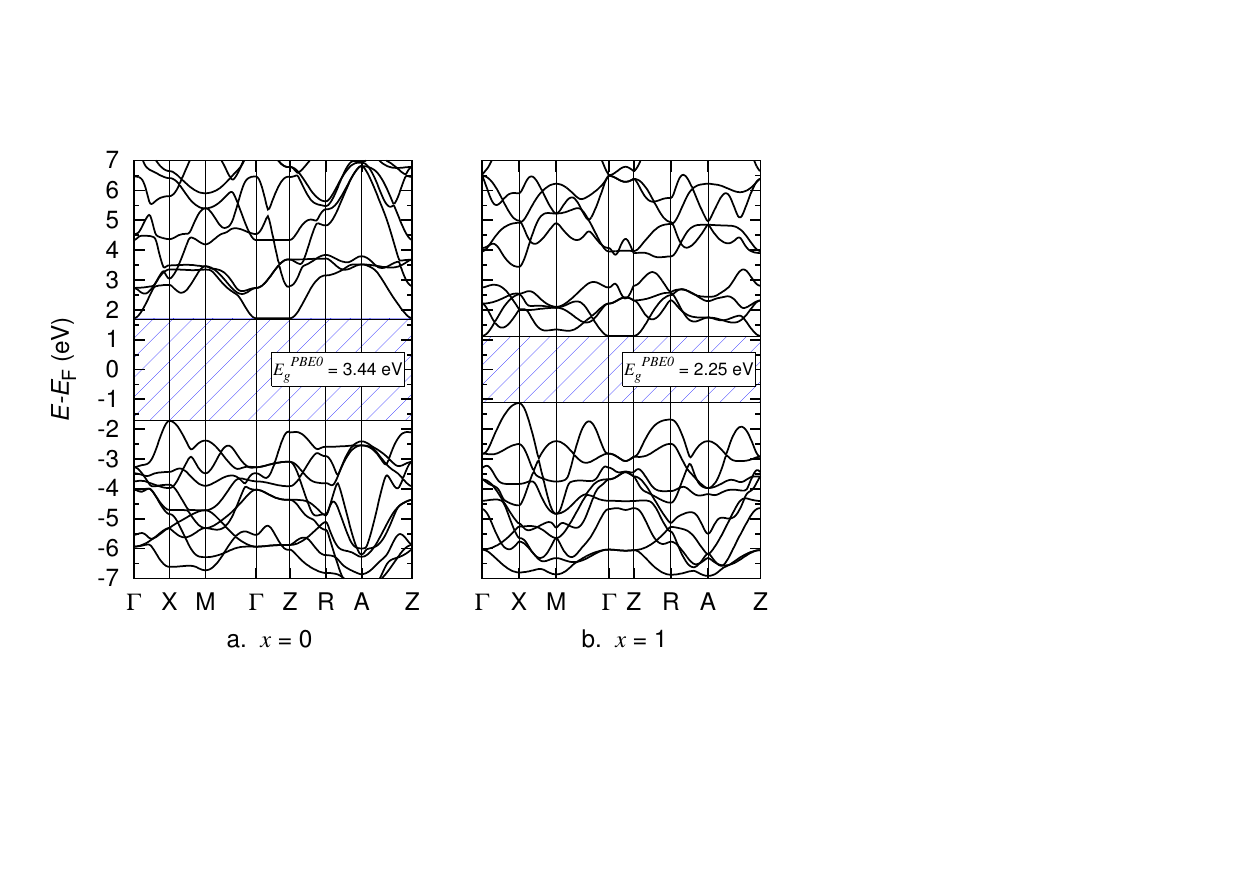}
\vspace{-110pt}
\caption{\label{justbands} LDA band structure with the conduction bands moved up and the valence bands moved down to portray the PBE0 band gap: (a.) PbTiO$_\text{3}$ ($x$ = 0) and (b.) PbTiO$_\text{2}$S  ($x$ = 1). }
\end{figure*}

As shown in Table \ref{elecres2}, the LDA band gap results are all in the infrared range for $x$ = 0.2 - 2.  Each gap is X-$\Gamma$ indirect, as seen experimentally for PTO.  Band structure diagrams for $x$ = 0 and $x$ = 1 are shown in Figure \ref{justbands}.  The narrowing of the band gap for $x$ = 1 relative to $x$ = 0 is evident.  The conduction bands are moved up and the valence bands are moved down to correct to the PBE0 values, due to the well-known LDA underestimation of $E_g$.  When PBE0 is applied to $x$ = 0.2 - 2, and $GW$ is applied to $x$ = 0.5, 1, and 2, the calculated gap magnitudes are well within the visible range.

The ranges of LDA band gap and polarization values for all relaxed ten-atom configurations for $x$~= 0.5, 1, and 2 are also included in Table \ref{elecres2}.   However, except for one configuration, the likelihood of achieving a different configuration, (and hence the associated electronic properties), than the high-symmetry one,  is very small, as they have relative energies greater than 0.27 eV/five-atoms higher than the respective ground state.  The one case, with S atoms sharing the edge of an octahedron in $x$ = 1, has a relative energy  0.05 eV/five-atoms higher  than the high symmetry ionic configuration and still has a significant LDA band gap (1.15 eV) and polarization (0.83 C/m$^\text{2}$).  Therefore, we expect that replacing O with S in these concentrations will lead to materials suitable for bulk photovoltaic use.

\begin{figure}[h]
\vspace{0pt}
\includegraphics[scale = 1.45]{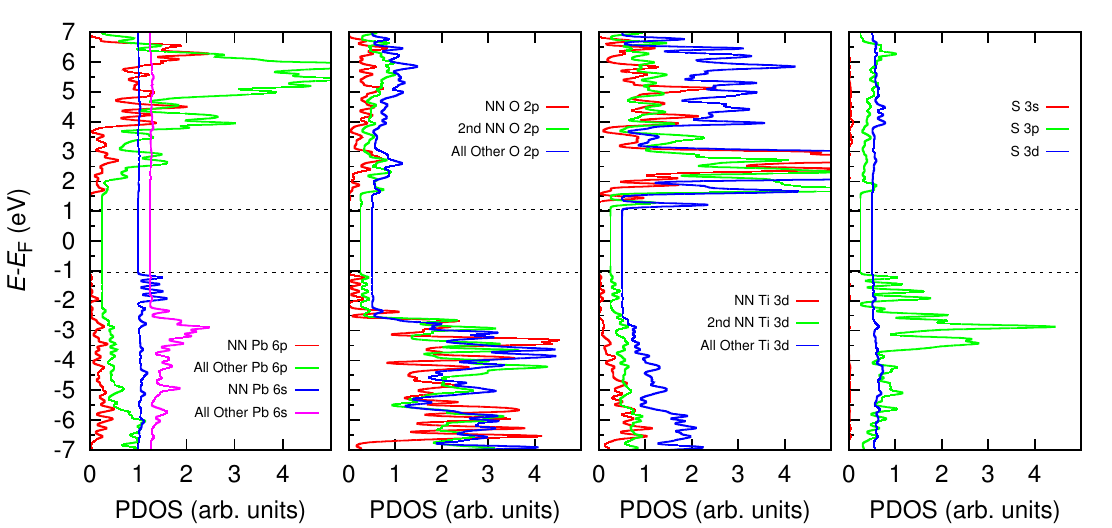}
\vspace{0pt}
\caption{\label{dos411} Orbital-projected density of states plots for PbTiO$_\text{3-$x$}$S$_\text{$x$}$,  $x$ = 0.25, with the conduction states moved up and the valence states moved down to portray the PBE0 band gap of 2.11 eV.  NN and 2$^\text{nd}$ NN stand for nearest neighbor and second nearest neighbor of the species with respect to S.}
\end{figure}

Atom-projected density of states calculations for $x$ = 0.20, 0.25, and 0.33 are remarkably similar.  A representative diagram is shown for $x$ = 0.25 in  Figure \ref{dos411}.   The densities of states show that the reduced band gap relative to PTO is a result of the higher energy S $3p$ orbitals relative to O $2p$ orbitals.     Note that the influence of S on the other atoms is  limited to its nearest neighbors.  Specifically, the top of the valence band has contributions from the $3d$ orbitals of the NN and second NN Ti atoms, the $6s$ and $6p$ orbitals of the NN Pb atom, and the $2p$ of the NN and second NN O atoms.

Bader charge analysis  can be used to estimate the ionic charges of atoms in molecules and compounds.  The Bader charge analysis results, from calculations performed on wavefunctions generated with the post-DFT PBE0 method, are shown in Table \ref{bader}.  These data show that as the sulfur concentration increases,  there is a monotonic decrease in the ionic charges of the cations and anions (with small exceptions) for the $x$ = 0, 0.5, 1, and 2 compositions, as the compounds become less ionic and more covalent.  This is due to the S being less electronegative than O.   Further, in $x$ = 0.5, all atomic species closer to S display ionicities smaller than their respective counterparts further away from S.   Bader charge analysis can also be used to confirm trends in band gaps of solutions.  In general, our Bader charge analysis results suggest that increasing the concentration of S will lead to compounds with smaller band gaps through reduced overall ionicity. However, on increasing S from $x$ = 0.5  to $x$ = 1, the PBE0 band gaps are 2.19 and 2.25 eV  but the overall ionicities are 3.54 and 3.46 .  An examination of PDOS alleviates this inconsistency and suggests a different correlation between ionicity and band gap.  In $x$ = 0.5, the PDOS indicates that electronic states of the NN Ti to S occupy the conduction band edge, while the other Ti (in $x$ = 0.5) has conduction band states that are approximately 0.015 eV higher in energy.  This NN Ti to S also has a smaller ionicity of 2.16 as compared to the other Ti (in $x$ = 0.5) which has an ionicity of 2.26.  More importantly, the ionicity of this NN Ti to S (in $x$ = 0.5) has a smaller ionicity than the Ti to S in $x$ =1 by 0.03 units.   Thus, these data indicate that the relevant correlation is between NN Ti ionicity and band gap, since the orbitals of these atoms set the edge of the conduction band.

\begin{table*}
\caption{\label{bader} Bader charge analysis results for $x$ = 0, 0.5, 1, and 2.  For $x$ = 0.5, two data entries are given per element:  the ionicity value for those atoms farther away from the S atoms is listed first. The summation of cation charges (or negative anion charges) per five atoms is represented by $\sum_{i}$C$_\text{i}$ = -$\sum_{i}$A$_\text{i}$.}
\begin{ruledtabular}
\begin{tabular}{c|cccc} 
      & \multicolumn{4}{c} {$x$} \\ \hline
 Species& 0       & 0.5 & 1& 2 \\     \hline

Pb	&	1.43	&	1.39, 1.26	&	1.27	&	1.13	\\
Ti	&	2.31	&	2.26, 2.16	&	2.19	&	2.05	\\
Equatorial O	&	-1.27	&	-1.27, -1.19	&	-1.25	&	-	\\
Equatorial S	&	-	&	-	&	-	&	-1.02	\\
Apical O	&	-1.19	&	-1.18	&	-	&	-1.14	\\
Apical S	&	-	&	-0.99	&	-0.97	&	-	\\ \hline
Charge sum & 3.73 &3.54 & 3.46 & 3.18 \\

\end{tabular}
\end{ruledtabular}
\end{table*}

\subsection {Formation energy results for replacing O with S in PbTiO$_\text{3}$}

Our literature review finds that no  PbTiO$_\text{3-$x$}$S$_\text{$x$}$ have been made.  Thus, in this section, we evaluate whether such synthesis is energetically feasible.  We consider replacing O with S via gaseous reagents.  While several experimenters have used H$_\text{2}$S and CS$_\text{2}$ to convert oxides to sulfides,\cite{Sato11p324,Cuya04p215,Saad10p2289,Lelieveld80p2223} Ishikawa \emph{et al.} have succeeded in replacing just the apical O of the TiO$_6$ octahedra with S to create ordered oxysulfides.\cite{Ishikawa04p2637}  We calculate the standard Gibbs free energy of reaction, $\Delta G^{\rm{0}}$, of creating PbTiO$_{3-x}$S$_x$ with $x$ = 0.2 - 3 in the energetically preferred configurations discussed above by comparing the sum of the $G^{\rm{0}}$ of the products to that
of the reactants for three different substitution scenarios:
\begin{eqnarray}
\mbox{PbTiO}_{3}(s) + \frac{x}{2}\mbox{S}_{2}(g) ~\rightarrow~ \mbox{PbTiO}_{3-x}\mbox{S}_{x}(s) + \frac{x}{2}\mbox{O}_{2}(g) \\
\mbox{PbTiO}_{3}(s) + x\mbox{H}_{2}\mbox{S}(g) ~\rightarrow~ \mbox{PbTiO}_{3-x}\mbox{S}_{x}(s) + x\mbox{H}_{2}\mbox{O}(g) \\
\mbox{PbTiO}_{3}(s) + \frac{x}{2}\mbox{CS}_{2}(g) ~\rightarrow~ \mbox{PbTiO}_{3-x}\mbox{S}_{x}(s) + \frac{x}{2}\mbox{CO}_{2}(g)
\end{eqnarray}
Here, standard state is defined as $p^0_{\rm{O_2}}$~= 1 bar =~0.987 atm.  In order to compute $\Delta G^{\rm{0}}$ we use:

\begin{eqnarray}
\Delta G^{\rm{0}}(T) &=& \ \  \ \ [E_{\rm{DFT, solid}} + F_{\rm{vib, solid}}(T) + H^{\rm{0}}_{\rm{gas}} -  T(S^{\rm{0}}_{\rm{gas}})]_{\rm{products}} \\& & - \ \ [E_{\rm{DFT, solid}} + F_{\rm{vib, solid}}(T) + H^{\rm{0}}_{\rm{gas}} - T(S^{\rm{0}}_{\rm{gas}})]_{\rm{reactants}} \notag\\ 
& &+ \ \ \Delta(pV) - T\Delta S^{\rm conf} \notag
\end{eqnarray}

\noindent where $p$ is pressure, $V$ is volume, and $S^{\rm{conf}}$ is configurational entropy.   In a constant pressure reaction,  at 0 K and one bar, the difference in $pV$ energy contribution of the solid products and reactants is on the order of 1$\times$10$^{-5}$ eV/5-atom.  At reaction temperatures, the volume difference between the products and reactants is not expected to change much, leaving $\Delta$($pV$) negligible, and it will not be considered further.   For $S^{\rm{conf}}$, only the solid product needs to be considered as the solid reactant is a pure compound and the gaseous species in both the reactant and product can be regulated to be predominantly the reactant gas, as was the case in the experimental works cited above in which flowing reactant gas was used.\cite{Lelieveld80p2223,Ishikawa04p2637} For calculation purposes, we assume an ideal solution, with no excess free energy of mixing.  

$G^{\rm{0}}_{\rm{solid}}$($T$) is described as the sum of solid state DFT total energy ($E_{\rm{DFT,solid}}$) and the harmonic vibrational Helmholtz free energy [$F_{\rm{vib,solid}}$($T$)].  The harmonic vibrational Helmholtz free energy is the sum of the harmonic vibrational internal energy and the product of temperature and the harmonic vibrational entropy:

\begin{eqnarray}
F_{\rm{vib, solid}}(T) = \sum_{ s=1}^{3N}\left\{ \frac{\hbar\omega_s}{2} + k_BT{\rm{ln}}\Big[1 - {\rm{exp}}\Big(\frac{-\hbar\omega_s}{k_BT}\Big)\Big]\right\}
\end{eqnarray}
\noindent where $N$ represents the number of atoms in the system, $\omega$$_s$ represents a $\Gamma$-point normal mode frequency, $k$$_B$ is the Boltzmann constant, and $T$ is temperature.

 For the gaseous species, the  $\Delta$$G^{\rm{0}}(T)$ values at finite temperatures are determined based on their $E_{\rm{DFT}}$ and the NIST-JANAF thermochemical tables of each species. We calculate molecular total energies by summing atomic energies obtained from spin polarized DFT calculations and the molecular atomization energies obtained from NIST.\cite{NISTweb}  Vibrational free energies of the gaseous species are determined using frequencies from the NIST-JANAF thermochemical tables.  As shown in Table \ref{thermo}, our calculations show that in the  temperature range 900 - 1300 K, replacing O with S in PTO is energetically favorable in CS$_\text{2}$($g$) and H$_\text{2}$S($g$) environments for $x$ = 0.2 - 0.5 at atmospheric pressure.  At 1300 K, S$_2$($g$) can be used to substitute O for S for $x$ = 0.2 and 0.25.

\begin{table*}
\caption{\label{thermo} {$\Delta G$$^\text{0}(T)$} (eV/5-atom PTO unit cell) calculations for various oxysulfide compositions formed by replacing O with S in PTO using the reactant indicated at 900, 1100, and 1300 K. }
\begin{ruledtabular}
\begin{tabular}{|c|ccc|ccc|ccc|} 

      & \multicolumn{3}{c|} {H$_\text{2}$S}   & \multicolumn{3}{c|} {CS$_\text{2}$} & \multicolumn{3}{c|}{S$_\text{2}$} \\
 $x$& 900 K& 1100 K &1300 K     & 900 K& 1100 K   & 1300 K& 900 K& 1100 K &1300 K \\  \hline
0.2	&	-0.26	&	-0.40	&	-0.54	&	-0.35	&	-0.49	&	-0.63	&	0.04	&	-0.10	&	-0.25\\
0.25	&	-0.20	&	-0.33	&	-0.47	&	-0.31	&	-0.44	&	-0.58	&	0.17	&	0.04	&	-0.10\\
0.33	&	-0.10	&	-0.21	&	-0.34	&	-0.25	&	-0.37	&	-0.49	&	0.39	&	0.27	&	0.14\\
0.5	&	0.11	&	0.01	&	-0.07	&	-0.12	&	-0.22	&	-0.31	&	0.85	&	0.75	&	0.65\\
1	&	0.74	&	0.74	&	0.74	&	0.29	&	0.28	&	0.28	&	2.24	&	2.21	&	2.19\\
2	&	1.08	&	1.01	&	0.95	&	0.17	&	0.10	&	0.02	&	4.03	&	3.94	&	3.85\\
3	&	1.81	&	1.74	&	1.66	&	0.46	&	0.36	&	0.26	&	6.24	&	6.13	&	6.00\\

\end{tabular}
\end{ruledtabular}
\end{table*}

\section{Conclusions}

We have shown that for the perovskite structure evaluated in ten-atom unit cells, the lowest energy state of PbTiO$_\text{3-$x$}$S$_\text{$x$}$ $x$ = 0.5, 1, and 2 is tetragonal with the minority species atoms located on apical sites of the octahedra.  The resulting structures for $x$ = 0.5 and 1, as well as for $x$ = 0.2, 0.25, and 0.33,  are tetragonal with $a$ lattice parameters within 2.5\% of the parent, PbTiO$_\text{3}$.  Our results also show that the use of CS$_\text{2}$($g$) and H$_\text{2}$S($g$) to replace O with S in PTO is a viable method to synthesize the compounds with $x$ = 0.2 - 0.5.
With respect to electronic properties, the polarization values of the $x$ = 0.2 - 1 materials are greater than that of the parent PTO and increase with increasing S concentration.    The band gaps of the $x$ = 0.2 - 2.0 systems  were evaluated by the post-DFT method of PBE0, and, for $x$ = 0.5, 1 and 2,  by the $GW$ method as well, and found to be in the visible range.   Thus, PbTiO$_\text{3-$x$}$S$_\text{$x$}$ $x$ = 0.2 - 2 are predicted to have significant polarization and low band gaps, and should be considered solar bulk photovoltaic material candidates.

\section{Acknowledgments}
HT and JWB were supported by the Office of Naval Research, under grant N00014-12-1-1033.   IG was supported by the National Science Foundation, under grant DMR11-24696. CWL was supported by  the Office of Naval Research, under grant N00014-11-1-0664.  JAB and MRS were supported by the Department of Energy Office of Basic Energy Sciences, under grant number DE-FG02-07ER46431.  AMR was supported by the Air Force Office of Scientific Research, Air Force Material Command, USAF, under grant FA9550-10-1-0248.  Computational support was provided by the HPCMO of the U.S. DoD and the NERSC center of the U.S. DoE.  The authors thank Prof. Graeme Henkelman for useful discussions on use of his group's Bader Analysis code.  The authors also thank Prof. Harold  Stokes  for useful communication concerning application of his group's FINDSYM code.

\end{document}